\documentclass[preprint,amsmath,amssymb,prl]{revtex4}

\usepackage{epsfig}

\def\be{\beta}

\def\rh{\rho}

\def\si{\sigma}

\def\ch{\chi}

\def\om{\omega}

\def\fr#1#2{{{#1} \over {#2}}}
\def\half{{\textstyle{1\over 2}}}

\def\frac#1#2{{\textstyle{{#1}\over {#2}}}}

\def\lsim{\mathrel{\rlap{\lower4pt\hbox{\hskip1pt$\sim$}}
    \raise1pt\hbox{$<$}}}
\def\gsim{\mathrel{\rlap{\lower4pt\hbox{\hskip1pt$\sim$}}
    \raise1pt\hbox{$>$}}}

\def\prt{\partial}

\def\etal{{\it et al.}}

\newcommand{\beq}{\begin{equation}}
\newcommand{\eeq}{\end{equation}}
\newcommand{\bea}{\begin{eqnarray}}
\newcommand{\eea}{\end{eqnarray}}
\newcommand{\bit}{\begin{itemize}}
\newcommand{\eit}{\end{itemize}}
\newcommand{\rf}[1]{(\ref{#1})}

\def\kb{\overline{k}{}}

\def\mbf#1{\boldsymbol #1}

\newcommand{\ke}{\overline{k}_{\mbox{\scriptsize eff}}}
\newcommand{\kei}[1]{(\kb_{\rm eff})_{#1}}

\begin{document}

\title{Search for Lorentz violation in short-range gravity}

\author{J.C.\ Long and V.\ Alan Kosteleck\'y}

\affiliation{
Physics Department, Indiana University,
Bloomington, IN 47405, U.S.A.}

\begin{abstract}

A search for sidereal variations in the force 
between two planar tungsten oscillators separated by about 80 $\mu$m 
sets the first experimental limits on Lorentz violation
involving quadratic couplings of the Riemann curvature,
consistent with no effect at the level of $10^{-7}$ m$^{2}$.

\end{abstract}

\maketitle

Local Lorentz invariance is a foundational component 
of General Relativity (GR),
which currently remains our most successful theory of gravity.
However,
GR is formulated as a classical theory,
and merging it with quantum physics in a consistent manner 
may well demand changes in its foundational structure. 
Even if local Lorentz invariance is exact 
in the underlying theory of quantum gravity,
spontaneous breaking of this symmetry may occur, 
leading to tiny observable effects
\cite{ksp}.
Experimental studies of Lorentz invariance
are therefore valuable as probes of the foundations of GR.

Short-range experiments 
are uniquely sensitive probes of gravity 
at scales below about a millimeter
and hence offer interesting opportunities 
to search for new physics beyond GR
\cite{review}.
The essence of short-range experiments is to measure the force
between two masses separated by a small distance.
To attain sensitivity at short range
without being overwhelmed by Newton forces at larger scales,
the test masses are typically scaled to that range. 
Experiments of this type are well suited,
for example,
to searching for deviations from the gravitational inverse-square law. 

To date,
most studies of local Lorentz invariance in gravity
are restricted to matter-gravity couplings
\cite{tables,cw}.
However,
recent theoretical work 
\cite{bkx}
using effective gravitational field theory
\cite{akgrav}
shows that quadratic curvature couplings involving Lorentz violation
lead to interesting new effects in short-range experiments
that could have escaped detection in conventional studies to date.
The Poisson equation for the Newton gravitational potential $U(\mbf r)$ 
generated by a source of mass density $\rh (\mbf r)$
acquires an extra perturbative term
with four spatial derivatives,
\beq
-\vec \nabla^2 U = 
4\pi G_N \rh 
+ (\kb_{\rm eff})_{jklm} \prt_j \prt_k \prt_l \prt_m U,
\label{fish}
\eeq
where $(\kb_{\rm eff})_{jklm}$ 
are effective coefficients with dimensions of squared length
that can be taken as constant on the scale of the solar system
\cite{sme}.
The extra term violates rotation symmetry and hence Lorentz invariance.
It is the general leading-order term in a natural perturbative expansion 
because a term with three derivatives is excluded by Newton's third law 
\cite{bkx}.
The presence of four derivatives implies corrections to the Newton force
that are inverse quartic and hence appear only at short range.
The rotation violation implies effects in the laboratory
depending on orientation 
and also on sidereal time due to the rotation of the Earth,
thereby ensuring that the resulting experimental signals are distinct 
from those associated with conventional 
Yukawa or inverse-power corrections.
The extra term offers a general description 
of dominant noncentral short-range corrections to Newton gravity
arising from an underlying unified theory.

Here,
we present new data acquired in March 2012
from a short-range experiment 
\cite{jcl02,jcl03,jcl14}
located in Bloomington, IN.
We use these data to search for sidereal variations
involving noncentral inverse-quartic corrections to Newton's law,
obtaining first constraints
on quadratic Lorentz-violating curvature couplings 
at the level of $10^{-7}$ m$^{2}$.
We also extend the analysis to incorporate the 2002 dataset
obtained with the apparatus located in Boulder, CO
\cite{jcl03}.
Note that existing searches for pure-gravity local Lorentz violation 
within this framework 
have been restricted to the context of a Lorentz-violating 
inverse-{\it square} law
\cite{2007Battat,2007MullerInterf,2009Chung,%
jcl10,2012Iorio,2013Bailey,2014Shao,bk}.
A few other short-range experiments
\cite{eotwash07,hust12,murata,irvine85}
may have potential sensitivity to the modifications \rf{fish},
while some experiments optimized for nonperturbative
corrections to Newton's law 
could conceivably be adjusted to study perturbative effects
\cite{iupui03,kapitulnik,geraci,tino14}.
Note also that constraints on forces with various inverse-power laws
have appeared in the literature 
\cite{inversesquare},
but only in the context of Lorentz-invariant effects.

\begin{figure}
\includegraphics[width=\hsize]{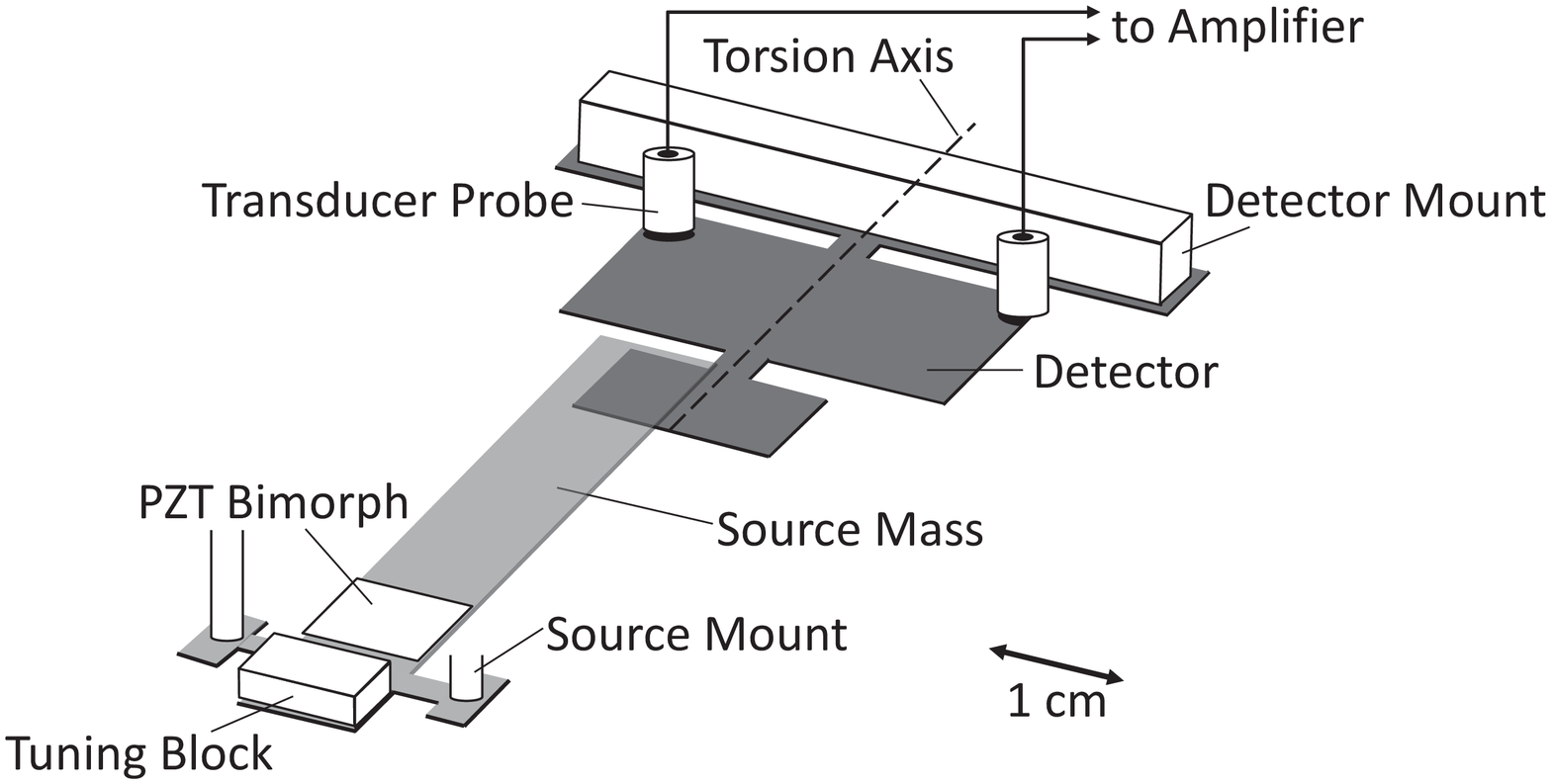}
\caption{
Schematic of the Indiana short-range experiment.}
\label{app}
\end{figure}

The design and operation of the experiment
is described elsewhere 
\cite{jcl02,jcl03,jcl14}.
Here, 
we summarize briefly the basic features. 
Each of the two test masses is a planar tungsten oscillator
of approximate thickness 250 $\mu$m,
separated by a gap of about 80~$\mu$m,
arranged as shown in Fig.\ \ref{app}. 
A stiff conducting shield 
is placed between them to suppress electrostatic and acoustic backgrounds.  
The planar geometry concentrates as much mass as possible 
at the scale of interest
while being nominally null with respect to inverse-square forces,
thereby suppressing the Newton background 
relative to new short-range effects. 
The force-sensitive `detector' mass is driven 
by the force-generating `source' mass at a resonance near 1 kHz.
Vibration isolation is a key requirement for this setup,
and operation at 1 kHz is chosen because at this frequency 
a comparatively simple passive vibration-isolation system can be used.
The entire apparatus is enclosed in a vacuum chamber 
and operated at $10^{-7}$ torr 
to minimize the acoustic coupling.  
Detector oscillations are read out via capacitive transducer probes 
coupled to a sensitive differential amplifier,
with the signal fed to a lock-in amplifier 
referenced by the same waveform used to drive the source mass.  
This design has proved effective in suppressing all background forces 
to the extent that only thermal noise is observed, 
arising from dissipation in the detector mass.  
The output of the lock-in amplifier constitutes the raw data.
These data are converted to force readings 
by comparison with the detector thermal noise, 
the scale of which is determined using the equipartition theorem 
\cite{jcl02}. 
Following data collection in 2002, 
this experiment set the strongest limits 
on unobserved forces of nature between 10 and 100 $\mu$m 
\cite{jcl03}.  
The apparatus has since been optimized to explore gaps below 50 $\mu$m,
and operation at the thermal noise limit has recently been demonstrated
\cite{jcl14}.

Measuring the coefficients $(\kb_{\rm eff})_{jklm}$ in Eq.\ \rf{fish} 
is the goal of the present analysis.
The coefficients are totally symmetric,
implying 15 independent observables for Lorentz violation.
Following standard convention,
we extract values of these observables in the canonical Sun-centered frame
\cite{tables,sunframe},
with $Z$ axis along the direction of the Earth's rotation
and $X$ axis pointing towards the vernal equinox.
As the Earth rotates,
the coefficients measured in the laboratory vary with sidereal time $T$.
The Earth's boost $\be_\oplus\simeq 10^{-4}$ can be neglected here. 
The transformation from the Sun-centered frame $(X,Y,Z)$
to the laboratory frame $(x,y,z)$ 
therefore involves a time-dependent rotation $R^{jJ}(T)$
\cite{bkx}
that depends on the Earth's sidereal frequency
$\om_\oplus\simeq 2\pi/(23{\rm ~h} ~56{\rm ~min})$ 
and the colatitude $\ch$ of the laboratory, 
which is 0.887 in Bloomington 
and 0.872 in Boulder.
The laboratory coefficients
$(\kb_{\rm eff})_{jklm} (T)$
are thus related to the coefficients
$(\kb_{\rm eff})_{JKLM}$
in the Sun-centered frame by
\beq
(\kb_{\rm eff})_{jklm} (T) 
= R^{jJ} R^{kK} R^{lL} R^{mM} (\kb_{\rm eff})_{JKLM}.
\label{rot}
\eeq

The cartesian components $g^j(\mbf r, T)$ 
of the modified gravitational acceleration 
at position $\mbf r$ and at sidereal time $T$
contain the conventional Newton acceleration 
along with an inverse-quartic correction term,
\bea
g^j (\mbf r, T) &=&  
-G_N \int d^3 r^\prime 
\rh (\mbf r^\prime )
\left( \fr{\widehat{R}{}^j}
{|\mbf r - \mbf {r^\prime}|^2}
+ \fr { \kb{}^j (\widehat{\mbf R}, T)} 
{|\mbf r - \mbf {r^\prime} |^4} \right).
\nonumber\\
\label{force}
\eea
Here,
$\widehat {\mbf R}=
(\mbf r - \mbf {r^\prime})/|\mbf r - \mbf {r^\prime} |$,
while
\bea
\kb^j (\widehat{\mbf R},T) &=& 
\frac{105}{2}(\ke)_{klmn}
\hat {R}^{j}\hat {R}^{k}\hat {R}^{l}\hat {R}^{m}\hat {R}^{n}
\nonumber\\
&&
-45(\ke)_{klmm} \hat{R}^{j} \hat{R}^{k} \hat{R}^{l}
+\frac{9}{2}(\ke)_{klkl}\hat {R}^{j}
\nonumber\\ && 
-30(\ke)_{jklm}\hat {R}^{k}\hat {R}^{l}\hat {R}^{m}
+18(\ke)_{jkll}\hat {R}^{k}
\qquad 
\label{tildekb}
\eea
controls the inverse-quartic force correction,
which varies with direction $\hat{\mbf R}$ and sidereal time $T$.
Note that the $T$ dependence is oscillatory
and includes components up to the fourth harmonic of $\om_\oplus$.

The detector is a constrained mechanical oscillator with distributed mass.
The modal amplitude at any point in the detector mass is
strongly dominated by vertical motion. 
This is particularly true near the thermal noise limit, 
where the amplitudes are of order 1 pm
\cite{jcl14}.
The experiment is thus sensitive predominantly 
to the $z$ component $F_p$ of the effective force 
at the location of the capacitive probe,
which can be written as
\beq
F_p (T) =
\fr{1}{d}\int_{D}d^3r~ \xi(\mbf r) F^z(\mbf r, T).
\label{for}
\eeq
Here, 
$\xi(\mbf r)$ is the detector mode-shape function,
which is the amplitude of the displacement of the detector at point $\mbf r$
when undergoing free oscillations in the relevant mode of interest,
and the displacement $d$ is the oscillation amplitude 
of the detector at the location of the probe.  
These quantities are derived from a finite-element model of the detector mass
and have the same arbitrary normalization.
The integration is taken over the volume $D$ of the detector 
over which the force is applied.

For the purposes of the present analysis,
Eq.\ \rf{for} is evaluated by Monte-Carlo integration,
using the $z$ component $F^z(\mbf r)$ of the force \rf{force}
expressed in terms of the coefficients $(\kb_{\rm eff})_{JKLM}$
in the Sun-centered frame
along with the geometrical parameters listed 
in Table II of Ref.\ \cite{jcl02}.
Note that the source amplitude for the 2012 dataset
was $22.2\pm 3.2$ $\mu$m
and the average gap was $77.5 \pm 20$ $\mu$m. 
The experiment is performed on resonance, 
so the Monte-Carlo algorithm computes 
the Fourier amplitude of Eq.\ \rf{for}
averaged over a complete cycle of the source-mass oscillation, 
taking into account the measured source-mass curvature and mode shape.  
The result can be expressed as a Fourier series in the sidereal time $T$,
\beq
F_p (T)=
\half C_0 +
\sum_{m=1}^{4}
S_{m\omega}\sin(m\omega_{\oplus}T)+C_{m\omega}\cos(m\omega_{\oplus}T).
\label{fourier}
\eeq
The Fourier amplitudes in this expression
are linear combinations of the coefficients $(\kb_{\rm eff})_{JKLM}$.
Their weights are functions 
of the source and detector mass geometry
and the laboratory colatitude.
Using approximately 500 million random pairs of points 
for each test mass suffices to resolve all harmonics.
Systematic errors from the dimensions and positions of the test masses
\cite{jcl02}
can be determined at this stage,
by computing the mean and standard deviation
of a population of Fourier amplitudes 
generated with a spread of geometries based on metrology errors. 
For the 2002 data,
the systematic error on the weights ranges 
from about 10\% to 75\%.
For the 2012 data,
it ranges from 15\% to 50\% 
on the most resolvable terms, 
while a few poorly resolved ones 
have systematic errors in excess of 100\%.  
Most of the systematic error 
is due to the uncertainty in the average gap, 
with a smaller contribution from the source amplitude.

All 15 independent components of $(\kb_{\rm eff})_{JKLM}$
appear in the Fourier series \rf{fourier},
although no single amplitude contains all of them.
The transformation \rf{rot} predicts some simple relations 
among the amplitudes, 
each of which is satisfied by the results of the numerical integration.
Performing the numerical integration for a hypothetical geometry
with an average gap an order of magnitude larger 
than the largest dimension of either mass
produces a result agreeing to within a few percent with 
the analytical expression for point masses of the same mass and separation.
This limiting case confirms that some contributions 
from $(\kb_{\rm eff})_{JKLM}$
are resolvable only due to the planar geometry.

\begin{figure}
\includegraphics[width=\hsize]{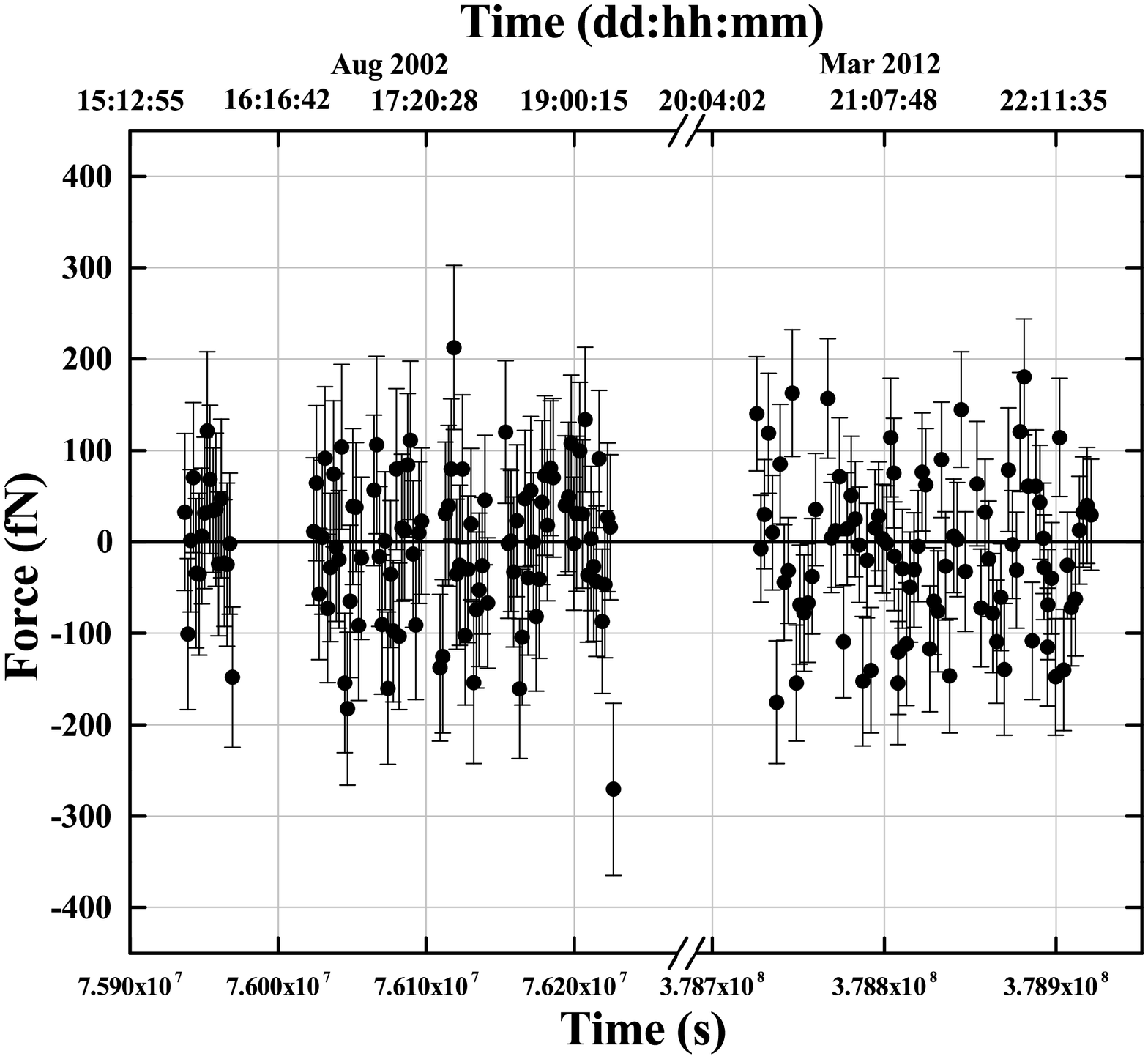}
\vskip -10pt
\caption{
Data from the Indiana short-range experiment.}
\label{fig:data}
\end{figure}

Figure \ref{fig:data} displays the force data 
acquired during the runs in 2012 and in 2002
as a function of the sidereal time $T$ 
measured in seconds from $T=0$,
which is taken to be the 2000 vernal equinox. 
The force data were collected at a 1 Hz rate
in 14.4-minute sets (2012 run) and in 12-minute sets (2002 run),
with comparable intervals between each set 
during which diagnostic data were taken 
to monitor the experiment for gain and frequency drifts.  
Each data point represents the mean of a 
14.4- or 12-minute set. 
Each error bar shown is the 1$\si$ standard deviation of the mean,
including both the statistical uncertainty 
and the systematic errors associated with the force calibration.
The 2002 force calibration and parameters are given 
by Eq.\ (2) and Table 1 of Ref.\ \cite{jcl02}.
The 2012 parameters are unchanged except that 
the mechanical quality factor was 22479$\pm$64, 
the resonance frequency was 1191.32$\pm$ 0.015 Hz, 
and the integrated mode shape was (6.0$\pm$0.6)$\times 10^{-11}$ m$^{5/2}$.
The calibration uncertainties for the 2002 and 2012 data
increase the errors by about 1\% and 2\%,
respectively.

Figure \ref{fig:data} represents a finite time series of force data 
with uneven time distribution.  
To analyze the data for Lorentz violation,
we adopt a well-established procedure
\cite{2009Chung}.
The ideal measure of each harmonic signal component 
is the corresponding Fourier amplitude in Eq.\ \rf{fourier}.
Each of these nine amplitudes,
$k=1,\ldots 9$, 
can be estimated by the discrete Fourier transform
$\tilde{d}_k = \fr{2}{N} \sum_{j}f(T_j)a_{k}(T_j)$,
where 
$N$ is the total number of force-data points 
plotted in Fig.\ \ref{fig:data}, 
$f(T_j)$ are the values of the force at each time $T_j$, 
and $a_{k}(T_j)$ 
is either $\sin(\om_k T_j)$ or $\cos(\om_k T_j)$ 
with $\om_k = m\om_\oplus$, $m=0,1,2,3,4$.
For this part of the analysis,
we treat the 2012 and 2002 results as separate datasets. 
The nine components $\tilde{d}_k$ extracted 
from the 2012 dataset and from the 2002 dataset 
are listed in the second and fourth columns of Table \ref{tab:ft}. 
The uncertainties are determined by propagating 
the errors of the time-series data in Fig.~\ref{fig:data}. 
The uncertainties can also be estimated 
by computing the Fourier transforms 
at several frequencies above and below the signal frequency
and calculating the root mean square of the values obtained.  
The former method is slightly more pessimistic
and is adopted here.

\renewcommand{\tabcolsep}{3pt}
\begin{table}
\begin{tabular}{lcccc}
\hline
\hline
\multicolumn{1}{c} {}& 
\multicolumn{1}{c} {2012 data}& 
\multicolumn{1}{c} {2012 data}& 
\multicolumn{1}{c} {2002 data}& 
\multicolumn{1}{c} {2002 data}\\ 
\multicolumn{1}{l}{Mode} & 
\multicolumn{1}{c}{$\tilde{d}_{k}$} & 
\multicolumn{1}{c}{$\tilde{D}_{k}$} &
\multicolumn{1}{c}{$\tilde{d}_{k}$} & 
\multicolumn{1}{c}{$\tilde{D}_{k}$}\\
\hline
$C_{0}$       & $-8.1\pm 5.0$  & $-3.1\pm 6.2$  & $-4.2\pm 7.8$   & $1.7\pm 19.1$\\
$S_{\omega}$  & $-0.7\pm 6.8$  & $-2.9\pm 8.7$  & $24.9\pm 9.6$   & $14.4\pm 22.9$\\
$C_{\omega}$  & $7.5\pm 7.3$   & $7.2\pm 7.8$   & $-2.2\pm 12.2$  & $-2.6\pm 11.5$\\
$S_{2\omega}$ & $-4.1\pm 7.1$  & $-10.1\pm 8.7$ & $-16.9\pm 12.0$ & $-4.3\pm 12.4$\\
$C_{2\omega}$ & $-9.4\pm 7.0$  & $-11.3\pm 7.6$ & $-0.8\pm 9.9$   & $-11.0\pm 26.4$\\
$S_{3\omega}$ & $-17.2\pm 7.1$ & $-18.9\pm 7.4$ & $33.5\pm 10.4$  & $30.8\pm 20.8$\\
$C_{3\omega}$ & $-11.8\pm 7.0$ & $-15.6\pm 8.9$ & $-19.2\pm 11.5$ & $-17.5\pm 12.6$\\
$S_{4\omega}$ & $-0.9\pm 7.1$  & $0.1\pm 7.6$   & $0.6\pm 11.3$   & $6.7\pm 13.8$\\
$C_{4\omega}$ & $3.4\pm 7.1$   & $-1.1\pm 8.1$  & $9.1\pm 10.7$   & $8.8\pm 21.7$\\
\hline
\hline
\end{tabular}
\caption{\label{tab:ft} 
Fourier transforms in fN units.}
\end{table}

For a finite time series, 
the Fourier components overlap. 
The overlap can be quantified by a correlation covariance matrix
cov$(a_k,a_{k^{\prime}})=
(2/N)\sum_{j}a_k(T_j)a_{k^{\prime}}(T_j)$.
The covariance matrix relates
the amplitudes $\tilde{D}_k$ for continuous data 
to the amplitudes $\tilde{d}_k$ for discrete data
according to
$\tilde{d}_k=
\sum_{k^{\prime}}{\rm cov}(a_k,a_{k^{\prime}}) \tilde{D}_{k^{\prime}}$.
The nine continuous amplitudes $\tilde{D}_{k}$
can be obtained by applying the inverse matrix cov$^{-1}$.
For the 2012 and 2002 datasets,
the results of this calculation are also displayed 
in the third and fifth columns of Table \ref{tab:ft}.
The $\tilde{D}_{k}$ can be taken to represent 
the measured values of the force components.
These values largely are consistent with zero within the quoted errors,
which include the small calibration systematics along 
with statistical errors.
The modes at $3\om$ appear to display resolved signals at this stage.
However, 
the associated coefficient weights are tiny,
so these force components become swamped by position systematics 
in the final results below.

\renewcommand{\tabcolsep}{3pt}
\begin{table}
\begin{tabular}{cccc}
\hline
\hline
\multicolumn{1}{l}{Coefficient}\quad & 
\multicolumn{1}{c}{2012 value}\quad & 
\multicolumn{1}{c}{2002 value}\quad & 
\multicolumn{1}{c}{Combined} \\
\multicolumn{1}{l}{} & 
\multicolumn{1}{c}{($10^{-7}$~m$^{2}$)}&
\multicolumn{1}{c}{($10^{-7}$~m$^{2}$)}& 
\multicolumn{1}{c}{($10^{-7}$~m$^{2}$)} \\
\hline
$\kei{XXXX}$ & $1.1\pm 3.2$& $0.5\pm 16.1$& $1.1\pm 3.1$\\
$\kei{YYYY}$ & $0.5\pm 3.2$& $-1.7\pm 16.2$& $0.4\pm 3.1$\\
$\kei{ZZZZ}$ & $0.6\pm 2.5$& $-0.7\pm 14.9$& $0.6\pm 2.5$\\
$\kei{XXXY}$ & $5.3\pm 19.5$& $2.5\pm 34.9$& $-3.4\pm 15.8$\\
$\kei{XXXZ}$ & $-9.5\pm 13.7$& $-5.9\pm 28.7$& $-8.1\pm 10.7$\\
$\kei{YYYX}$ & $-5.7\pm 19.5$& $-1.0\pm 35.1$& $-4.4\pm 15.8$\\
$\kei{YYYZ}$ & $7.3\pm 12.2$& $-31.7\pm 44.8$& $4.6\pm 9.6$\\
$\kei{ZZZX}$ & $-6.6\pm 21.5$& $-3.5\pm 45.7$& $-5.4\pm 16.6$\\
$\kei{ZZZY}$ & $7.4\pm 23.3$& $-12.6\pm 45.8$& $3.4\pm 17.8$\\
$\kei{XXYY}$ & $0.4\pm 1.6$& $-0.5\pm 8.5$& $0.4\pm 1.5$\\
$\kei{XXZZ}$ & $0.2\pm 1.6$& $-0.5\pm 9.1$& $0.2\pm 1.6$\\
$\kei{YYZZ}$ & $0.6\pm 1.6$& $-0.3\pm 9.1$& $0.5\pm 1.6$\\
$\kei{XXYZ}$ & $-3.6\pm 5.7$& $16.2\pm 25.0$& $-2.7\pm 5.5$\\
$\kei{YYXZ}$ & $4.7\pm 7.2$& $7.5\pm 17.2$& $5.0\pm 6.6$\\
$\kei{ZZXY}$ & $-4.0\pm 6.5$& $-0.4\pm 2.1$& $-0.7\pm 1.9$\\
\hline
\hline
\end{tabular}
\caption{\label{tab:limits}
Coefficient values (2$\sigma$) 
from the 2012, 2002, and combined datasets,
with all other coefficients vanishing.}
\end{table}

Individual measurements of the independent components
of $(\kb_{\rm eff})_{JKLM}$
can be extracted from a global probability distribution 
formed using the values 
of the nine continuous amplitudes $\tilde{D}_{k}$ and their errors. 
Each measured amplitude can be assigned
a corresponding probability distribution 
$p_k = p_k((\kb_{\rm eff})_{JKLM})$
that is a function 
of the 15 independent components of $(\kb_{\rm eff})_{JKLM}$.
The $p_k$ are assumed to be gaussian 
with means $\mu_k$ and standard deviations $\si_k$.
The global probability distribution 
$P = P((\kb_{\rm eff})_{JKLM})$
of interest is then the product of the individual $p_k$,
taking the form
\beq
P = P_0 \exp\left[ -\sum_{k=1}^{9}
\fr {(\tilde{D}_k - \mu_k)^2} {2\si^2_k}
\right].
\label{Pk}
\eeq
In this expression, 
$P_0$ is an arbitrary normalization.  
The predicted signal 
$\mu_k = \mu_k((\kb_{\rm eff})_{JKLM})$
for the $k$th amplitude
is determined from Eqs.\ \rf{for} and \rf{fourier},
and the variance $\si_k^2$ includes all statistical and systematic errors.

An independent measurement of any one chosen component 
of $(\kb_{\rm eff})_{JKLM}$
can in principle be obtained 
by integrating the global probability distribution $P$
over all other components. 
The result is a distribution
involving the chosen component with a single mean and standard deviation,
which constitute the estimated component measurement and its error.
However,
the 2012 dataset alone contains only nine signal components,
which is insufficient to constrain independently
each of the 15 degrees of freedom in $(\kb_{\rm eff})_{JKLM}$.
Following standard practice in the field
\cite{tables},
we can obtain maximum-sensitivity constraints
on each component of $(\kb_{\rm eff})_{JKLM}$ in turn
by integrating the global probability distribution
with the other 14 degrees of freedom set to zero.
The resulting measurements and 2$\si$ errors
on each independent component of $(\kb_{\rm eff})_{JKLM}$
are displayed in the first two columns of Table \ref{tab:limits}. 
Note that the first column reveals our choice for 
the 15 independent components of $(\kb_{\rm eff})_{JKLM}$.
Note also that the sensitivity of the apparatus 
to the coefficients $(\kb_{\rm eff})_{JKLM}$ 
can be crudely estimated as 
the ratio of the thermal-noise force at the location of the probe
($\sim$10 fN)
to the scale ($\sim$$10~\mu$N/m$^{2}$)
of the amplitudes in the Fourier series \rf{fourier},
multiplied by a suppression factor of order $10^{-2}$
because the dominant contribution to the noncentral force
in a parallel-plate geometry arises from edge effects 
\cite{hust15}.
This estimate matches the size of the values 
in the second column of Table \ref{tab:limits}.

The third column of Table \ref{tab:limits} 
displays the values for the coefficients $(\kb_{\rm eff})_{JKLM}$ 
obtained from a comparable analysis of the 2002 dataset.
These 2002 results are about a factor of five
less sensitive than the 2012 data,
a feature that can be traced to the  
larger average gap between the source and detector masses
and the smaller source-mass amplitude in the 2002 experiment.  
The final column of Table \ref{tab:limits} 
presents the measured values
of each independent component taken in turn
that are obtained from analyzing the combined datasets.

The contents of Table \ref{tab:limits} 
represent the first measurements 
of noncentral inverse-quartic corrections to Newton gravity
and hence of quadratic curvature couplings violating local Lorentz invariance.
The inverse-quartic dependence implies the corrections
are perturbative at squared distances greater than the coefficient values.
For example, 
the perturbative effects at the apparatus scale 
are roughly comparable to the Newton force,
while on the macroscopic scale of the laboratory 
the attained constraints exclude noncentral forces
at about parts in ten million.
An alternative perspective can be obtained by comparing
the length dimension associated with the coefficients 
$(\kb_{\rm eff})_{JKLM}$
to the various scales set by the Compton wavelengths
of elementary particles. 
The experiment here probes modifications governed 
to within about an order of magnitude
of the scale of the neutrino Compton wavelength.
Effects at the scales of Compton wavelengths 
of other particles would be smaller,
reflecting the possibility that comparatively large
`countershaded' Lorentz violation
remains a viable possibility
\cite{kt}. 

The results reported here set a benchmark for future efforts.
For example,
upgrading the apparatus used 
by improving the test-mass and shield flatness
could reduce the average gap by a factor of two,
and refining the test-mass metrology 
could reduce the uncertainty in the average gap by a factor of four.
Simulations suggest these improvements would increase the overall sensitivity
by more than an order of magnitude in the absence of new systematics.
With several months of run time, 
the statistical error bars could be reduced
by about another order of magnitude.
Moreover,
other experimental groups also have the capability
of improving substantially over the results in the present work 
\cite{bkx}.
For example,
the HUST experiment has recently reported sensitivities
to the coefficients $(\kb_{\rm eff})_{JKLM}$
surpassing those reported here
\cite{hust15}.
Overall,
the prospects for improved future short-range searches 
for Lorentz violation are excellent.

\medskip

We thank 
Q.\ Bailey, R.\ Decca, and R.\ Xu for discussions,
S.\ Kelly for collection of the 2012 data,
and D.\ Bennett and W.\ Jensen
for work on the Monte-Carlo code
in earlier incarnations of this experiment.
We are grateful to  
C.-G.\ Shao, Y.-J.\ Tan, W.-H.\ Tan, S.-Q.\ Yang, J.\ Luo, and M.E.\ Tobar
for drawing our attention to an issue with the Monte-Carlo code
used in the original version of this work.
The 2012 data were taken
at the Indiana University Center for the Exploration of Energy and Matter.
This work was supported in part 
by the National Science Foundation 
under grant number PHY-1207656,
by the Department of Energy
under grant number {DE}-SC0010120,
and by the Indiana University Center for Spacetime Symmetries.

\end{document}